\documentclass[twocolumn,showpacs,superscriptaddress,preprintnumbers,amsmath,amssymb,aps,pre,10pt]{revtex4-1}

\usepackage{graphicx}% Include figure files
\usepackage{color}
\usepackage{dcolumn}% Align table columns on decimal point
\usepackage{bm}% bold math

\begin{document}

%\preprint{APS/123-QED}

\title{Lattice cluster theory for polymer melts with specific interactions}

\author{Wen-Sheng Xu}
\email{wsxu@uchicago.edu}
\affiliation{James Franck Institute, The University of Chicago, Chicago, Illinois 60637, USA}

\author{Karl F. Freed}
\email{freed@uchicago.edu}
\affiliation{James Franck Institute, The University of Chicago, Chicago, Illinois 60637, USA}
\affiliation{Department of Chemistry, The University of Chicago, Chicago, Illinois 60637, USA}

\date{\today}% It is always \today, today, but any date may be explicitly specified

\begin{abstract}
Despite the long-recognized fact that chemical structure and specific interactions greatly influence the thermodynamic properties of polymer systems, a predictive molecular theory that enables systematically addressing the role of chemical structure and specific interactions has been slow to develop even for polymer melts. While the lattice cluster theory (LCT) provides a powerful vehicle for understanding the influence of various molecular factors, such as monomer structure, on the thermodynamic properties of polymer melts and blends, the application of the LCT has heretofore been limited to the use of the simplest polymer model in which all united atom groups within the monomers of a species interact with a common monomer averaged van der Waals energy. Thus, the description of a compressible polymer melt involves a single van der Waals energy. As a first step towards developing more realistic descriptions to aid in the analysis of experimental data and the design of new materials, the LCT is extended here to treat models of polymer melts in which the backbone and side groups have different interaction strengths, so three energy parameters are present, namely, backbone-backbone, side group-side group, and backbone-side group interaction energies. Because of the great algebraic complexity of this extension, we retain maximal simplicity within this class of models by further specializing this initial study to models of polymer melts comprising chains with poly($n$-$\alpha$-olefin) structures where only the end segments on the side chains may have different, specific van der Waals interaction energies with the other united atom groups. An analytical expression for the LCT Helmholtz free energy is derived for the new model. Illustrative calculations are presented to demonstrate the degree to which the thermodynamic properties of polymer melts can be controlled by specific interactions. 
\end{abstract}

%\pacs{64.70.pj, 83.80.Sg, 05.70.-a, 05.50.+q}

%\pacs{Valid PACS appear here}% PACS, the Physics and Astronomy Classification Scheme.
%\keywords{Suggested keywords}%Use showkeys class option if keyword display desired

\maketitle

\section{Introduction}

The development of reliable approaches for predicting the thermodynamic properties of polymer systems is crucial for the rational design of polymer materials. Such approaches should readily translate microscopic information into predictions for macroscopic properties. Although fully atomistic simulations often generate useful microscopic information, a unified framework for understanding the influence of molecular factors on the thermodynamic properties of polymer systems cannot be readily extracted from these simulations. The lattice cluster theory (LCT) of Freed and coworkers~\cite{JCP_87_7272, Mac_24_5076, ACP_103_335, APS_183_63} represents a class of coarse grained models that retain the essential features (e.g., bond connectivity, cohesive energy, chain stiffness, etc.) of molecular structure and interactions in polymer fluids while enjoying analytic tractability and computational simplicity. Therefore, the LCT systematically addresses the role of various molecular characteristics in determining the thermodynamic properties of polymer systems.

The LCT builds on the long-recognized mathematical simplification to the description of polymer systems conferred by lattice models whose use enables predicting new phenomena and explaining existing data. Classic Flory-Huggins (FH) theory~\cite{JCP_9_660, JCP_9_440, JCP_46_151} successfully explains the fact that long polymer chains in the liquid state generally tend to be immiscible, but it completely neglects the role of monomer structure and chain architecture in determining the thermodynamic properties of polymer systems. This significant deficiency is rectified by the LCT, which employs an extended lattice model where monomers possess explicit structures that extend over several lattice sites and preserve specific bonding patterns. This essential feature of the LCT provides a substantial improvement over the traditional lattice theories and enables studying the influence of monomer size, shape  and energetic disparities on the thermodynamic properties of polymer melts and blends. Moreover, the LCT includes the description of the short-range correlations imparted by nearest neighbor van der Waals interactions and by chain semiflexibility. Not surprisingly, the LCT provides microscopic explanations for a variety of previously enigmatic observations~\cite{JCP_87_7342, Mac_22_4048, JCP_97_2105, JCP_108_7881, Mac_24_5112, Mac_29_7826, Mac_29_8960, Mac_30_7279} and offers a valuable vehicle for predicting interesting new phenomena~\cite{Mac_26_213, Mac_33_5592, Mac_33_9777, Mac_24_5096, Mac_28_6625, PRL_88_095503, JCP_116_9983}. The successful applications of the LCT motivate further improving the theory in the present work.

One simplified assumption common to all previous applications of the LCT is the approximation of assigning an identical van der Waals interaction energy to all pairs of united atom groups within an individual polymer species (see recent reviews~\cite{APS_183_63, ACR_44_194, ACP_137_125} and the references therein). Although this monomer averaged interaction model suffices in establishing qualitative trends observed in real polymers, the simplified model clearly limits scope of assessing how the energetic heterogeneity within monomers can be used as an additional variable for controlling the thermodynamic properties of polymer systems. In fact, the differences in interaction energies are significant even for very similar united atom groups; e.g., typical Lennard-Jones energy parameters for $(\text{CH}_3)-(\text{CH}_3)$ and $(\text{CH}_2)-(\text{CH}_2)$ united atom group interactions are $90$ K and $70$ K, respectively~\cite{JCP_103_7156, JACS_103_335, JCP_93_4290}. Thus, it is quite natural to extend the LCT to models of polymers with specific interactions. 

The present paper is designed as a first step in the development of more realistic models to aid experimentalists in the interpretation of data and the design of new materials. The treatment here focuses on the general methods for developing these more detailed descriptions, with specific application to a model of a melt of semiflexible polymers with the structure of poly($n$-$\alpha$-olefin) chains in which only the united atom groups residing at the ends of the side chains are assigned different nearest neighbor van der Waals interaction energies. We provide a detailed description of the derivation of the analytic expression for the LCT Helmholtz free energy that enables evaluating all other thermodynamic quantities. Given the lengthiness of the derivation, only a few illustrative examples are provided for the influence of specific interactions on thermodynamic properties, while separate papers will present a detailed examination of the influence of molecular details on the thermodynamic properties of polymer melts, with particular emphasis on investigating polymer glass formation using a combination of the current extension of the LCT with the Adam-Gibbs theory~\cite{JCP_43_139}.  

Section II details the model of polymer melts with three different interaction energies, followed in Section III by a brief review of some necessary technical details from the previous developments of the LCT. Appendix A summarizes the detailed expressions for the Helmholtz free energy for the model melts with monomer averaged interactions for comparison with the current extension of the LCT in Section IV, with Appendices B and C providing the explicit analytic expressions for the Helmholtz free energy for the model of a melt with specific interactions. Section V presents illustrative calculations for the influence of specific interactions on the thermodynamic properties of polymer melts. 

\section{Model description}

\begin{figure}[tb]
 \centering
 \includegraphics[angle=0,width=0.45\textwidth]{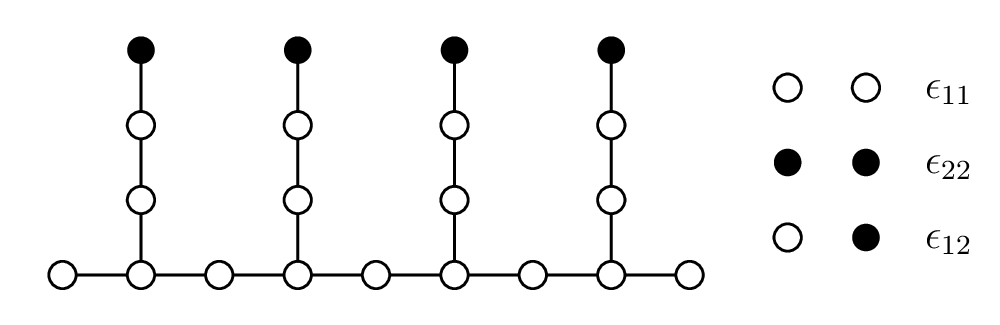}
 \caption{Schematic illustration of the lattice model of the polymer melts with specific interactions considered here. Circles designate united atom groups, while lines represent the bonds between united atom groups. The monomers (repeat units) are endowed with $n$-$\alpha$-olefin structures. The example depicted contains $N_s=3$ united atom groups in each side chain and $N_c=4$ repeat units. The united atom groups residing at the ends of each side chain have specific interactions, i.e., different nearest neighbor van der Waals interaction energies $\epsilon_{22}$ with each other and $\epsilon_{12}$ with all other united atom groups. Therefore, the model prescribes three different van der Waals interaction energies ($\epsilon_{11}$, $\epsilon_{22}$ and $\epsilon_{12}$).}
\end{figure}

The extended lattice model of polymers considers monomers that are endowed with $n$-$\alpha$-olefin structures. The individual monomers are thus represented in terms of a set of united atom groups that each occupies a single lattice site. As in previous work, the theory is developed using a $d$-dimensional hypercubic lattice with $N_l$ lattice sites, each with $z=2d$ nearest neighbors. The system is treated as a one-component compressible melt, consisting of $m$ monodisperse chains with $M$ united atom groups per chain, producing the volume fraction of the polymer chains as $\phi=mM/N_l$. Specializing to the present model depicted in Fig. 1, $M$ can be expressed as $M=N_c(N_s+2)+1$, where the backbone chain length $N_c$ and the side chain length $N_s$ are the numbers of monomers per chain and united atom groups per side chain. By construction, each side chain has at least one united atom group. 

The model in Fig. 1 is designed to investigate the importance of varying the chemical species of the side groups in real polymers. The united atom groups residing at the ends of the side chains (represented as solid circles in Fig. 1 and called e-groups in the following) differ from those lying on other positions (depicted by open circles in Fig. 1 and termed n-groups in the following), since these side chains' end segments have different nearest neighbor van der Waals interaction energies among them and with other n-groups. Therefore, each chain contains two types of united atom groups with $M_n=N_c(N_s+1)+1$ n-groups (species 1) and $M_e=N_c$ e-groups (species 2). We define the fractions of n-groups and of e-groups in a single chain as $f_n=M_n/M$ and $f_e=M_e/M$, respectively. The van der Waals energies $\epsilon_{11}$ and $\epsilon_{22}$ are assigned to the interaction of two nearest neighbor n-groups and two nearest neighbor e-groups, while the energy $\epsilon_{12}$ describes the interaction strength between a n-group and an e-group. These systems are thus termed polymer melts with specific interactions to distinguish them from polymer melts with monomer averaged interactions, where a single van der Waals energy specifies the interactions of the system. The models of melts with specific interactions, of course, reduce to that with monomer averaged interactions when all nearest neighbor van der Waals interaction energies become identical, i.e., $\epsilon_{11}=\epsilon_{22}=\epsilon_{12}$. 

Similar theoretical methods enable treating model polymers where all (or some) of the united atom groups residing in the side chains are species 2 (i.e., all side groups become solid circles in Fig. 1). Thus, sufficient details are presented to guide the treatment of these other models. Particular applications will be presented elsewhere for the simplest case where each side chain contains only one united atom group, i.e., the polymer chains formally have the structure of poly(propylene) (PP) because this choice involves the minimal number of energetic parameters as described below.

\section{General formulation of the lattice cluster theory}

This section begins with a brief sketch of the derivation of  the LCT partition function for a melt of semiflexible lattice polymers with structured monomers. The derivation of the general expression for the partition function follows the lines of earlier works but proceeds differently by introducing each additional feature in the model by constraints which are then converted into a schematic diagrammatic representation of the cluster expansion for the partition function. This constraint oriented review provides necessary notation and concepts for the extension of the theory to describe ever more realistic models and, in particular, models of melts with specific interactions. Detailed expressions for the Helmholtz free energy of the model melts with monomer averaged interactions are provided in Appendix A with some corrections to previous work~\cite{ACP_103_335}.

\subsection{LCT partition function and its cluster expansion}

The LCT enables deriving a systematic analytical approximation for the Helmholtz free energy, from which all other thermodynamic quantities can be calculated analytically. The LCT calculations begin by evaluating the partition function for the extended lattice model for a system of interacting polymers with structured monomers. To account for a variety of important characteristics of real polymers, the expression for the LCT partition function is constructed by imposing several constraints associated with the presence of excluded volume interactions, chain connectivity, nearest neighbor van der Waals interactions, semiflexibility, and monomer structures. We briefly review the nature of these constraints and their treatment in the LCT. Further details can be found in Ref.~\cite{ACP_103_335}. 

Excluded volume constraints account for the hard-core repulsions that exist in all systems at short interatomic separations. The Boltzmann factors for these short-ranged, non-bonded constraints translate in the lattice model to the condition that no two united atom groups can occupy the same lattice site. Thus, the partition function incorporates the excluded volume constraints by using a restricted sum over all positions of every united atom group. 

The treatment of the constraints due to chain connectivity simplifies in the lattice model to the condition that the vector $\mathbf{r}_{i}-\mathbf{r}_{i+1}$ connecting united atom groups $i$ and $i+1$ is one of the $z$ possible lattice vectors $\mathbf{a}_{\mu}$ ($\mu=1, ..., z$) between nearest neighbor sites on the lattice. This simplification can be expressed as 
\begin{equation}
\sum_{\mu_{i}=1}^{z}\delta_{\text{bond}}(\mathbf{r}_{i},\mathbf{r}_{i+1}+\mathbf{a}_{\mu_{i}}),
\end{equation}
where $\delta_{\text{bond}}$ is the Kronecker delta function specifying the presence or absence of a bond ($\delta_{\text{bond}}=1$ if united atom groups $i$ and $i+1$ are nearest neighbors, and $\delta_{\text{bond}}=0$ otherwise). The sum over $\mu$ appears because the bond may lie along any of $z$ possible directions.

Real polymers must have longer-range attractive interactions in order to attain liquidlike densities at atmospheric pressure. These longer-range interactions are modeled in the LCT with nearest neighbor attractive van der Waals energies $\epsilon_{ij}$, and the interactions are then approximated using a cluster expansion in terms of Mayer $f$ functions. For this purpose, the interactions between each pair of interacting united atom groups $i$ and $j$ enter into the expression for the partition function with the following factor,
\begin{equation}
1+f_{\text{pair}}\sum_{\mu_{i}=1}^{z}\delta_{\text{pair}}(\mathbf{r}_{i},\mathbf{r}_{j}+\mathbf{a}_{\mu_{i}}).
\end{equation}
The Mayer $f$ function is defined as $f_{\text{pair}}=f_{ij}=\exp(\beta\epsilon_{ij})-1$ with $\beta=k_BT$, where $k_B$ is Boltzmann's constant and $T$ is the absolute temperature, and $\delta_{\text{pair}}=1$ when the united atom groups $i$ and $j$ are nearest neighbors even if they are not bonded, and $\delta_{\text{pair}}=0$ otherwise. The van der Waals energy $\epsilon_{ij}$ is written as positive in the LCT, but with the understanding that $\epsilon_{ij}$ describes the net attractive interactions. Therefore, the interactions between united atom groups become more attractive for larger $\epsilon_{ij}$. Since the original LCT takes $\epsilon_{ij}$ as independent of united atom groups $i$ and $j$ for a melt, the previously derived partition function applies to the models of melts with monomer averaged interactions, while the extension to allow variable $\epsilon_{ij}$ is considered in the present work for application to the models of melts with specific interactions.

To model the presence of conformational energies that are responsible for a stiffening of chains upon cooling, a chain semiflexibility constraint is introduced into the LCT partition function for a pair of consecutive bonds as
\begin{equation}
E_{\text{bend}}+(1-E_{\text{bend}})\delta_{\text{bend}}(\mu_{i},\mu_{i+1}),
\end{equation}
where $E_{\text{bend}}=\exp(-\beta E_b)$ with $E_b$ called the bending energy, and $\delta_{\text{bend}}$ depends on the conformation of the two sequential bonds ($\delta_{\text{bend}}=1$ if the two bonds are collinear and $\delta_{\text{bend}}=0$ otherwise). The bending energy $E_b$ has been introduced as the energy difference between \textit{trans} and \textit{gauche} conformations for a pair of consecutive bonds. Specifically, the \textit{trans} conformation corresponds to consecutive parallel bonds and is ascribed a vanishing bending energy, while $E_b$ is prescribed to a \textit{gauche} pair of sequential bonds lying along orthogonal directions. By definition, the bending energy $E_b$ is positive, and chains are fully flexible for $E_b=0$, while they become completely rigid in the limit $E_b\rightarrow\infty$.

Finally, the LCT incorporates the bonding constraints associated with monomer structure, which is described with the aid of the concept of \textit{subchains} in which the full structured polymer chain is decomposed into a minimal set of linear chains, called subchains, that share no bonds in common but that may share united atom groups at junction points. For example, a PP chain contains $L$ = $N_c+1$ subchains, where $N_c$ is the polymerization index, since the backbone constitutes one subchain, while each side chain bond forms a separate subchain. The subchain is also an important concept for semiflexible polymers and special attention must be devoted to prescribing the bending energy factor to structured chains (see more explanations in Ref. ~\cite{ACP_103_335, JCP_119_5730}). Since these constraints depend on the monomer structure, they are most conveniently applied in the diagrammatic representation of contributions to the partition function. Hence, the monomer structure constraints do not appear in the partition function in the form of an explicit constraint.

The exact expression for the partition function $W$ is conveniently represented using the schematic form of Eq. (17) in Ref. \cite{ACP_103_335} that emphasizes the physical nature of individual constraints,
\begin{eqnarray}
W=&&\frac{1}{2^mm!}\sum_{\{\mathbf{r}\}}{\textstyle{'}}
\left\{ \prod_{\text{bond}}\left( \sum_{\mu_{\text{bond}=1}}^{z}\delta_{\text{bond}} \right) \right. \nonumber\\
&&
\times \prod_{\text{pair}} \left( 1+f_{\text{pair}}\sum_{\mu_{\text{pair}}=1}^{z}\delta_{\text{pair}} \right)
\nonumber\\
&&
\times \left. \prod_{\text{bend}}\left[ E_{\text{bend}}+(1-E_{\text{bend}})\delta_{\text{bend}} \right] \right\},
\end{eqnarray}
where $m$ is the number of the chains, the factors of $2^{-m}$ and $1/m!$ account for the indistinguishability of the two chain ends and of the chains themselves, the symbol $\sum'_{\{\mathbf{r}\}}$ represents a restricted sum over the positions of all united atom groups subject to the constraint prohibiting multiple occupancy of any lattice site, the bond product includes all bonds between united atom groups, the pair product runs over all pairs of united atom groups that reside on neighboring lattice sites, and the product labeled bend runs over all pairs of consecutive bonds within each subchain.

The formally exact expression for the partition function $W$ is evaluated using a systematic cluster expansion in which the zeroth-order mean-field approximation ($W^{mf}$ in Eq. (19) of Ref.~\cite{ACP_103_335}) is essentially equivalent to Flory-Huggins theory, as explained in more detail in Ref.~\cite{ACP_103_335}. This approximation enables re-expressing the partition function $W$ as the product of a zeroth-order mean-field contribution and corrections, i.e., $W=W^{mf}(1+corrections)$. The term $corrections$ can be converted into a systematic cluster expansion by first noting the fact that each Kronecker delta in Eq. (4) can be re-expressed in the schematic form
\begin{equation}
\delta=A+(\delta-A)=A\left( 1+\frac{\delta-A}{A} \right),
\end{equation}
where the term $A$ represents an average of the contribution from $\delta$ and thus defines a factor appearing in the zeroth-order mean-field term $W^{mf}$. For example, the bond term $\sum_{\mu_{\text{bond}=1}}^{z}\delta_{\text{bond}}$ in $W$ can be rewritten exactly as  
\begin{equation}
\sum_{\mu_{\text{bond}=1}}^{z}\delta_{\text{bond}}=zA_{\text{bond}}(1+X_{\text{bond}}),
\end{equation}
where $zA_{\text{bond}}=z/N_l$ represents a simple zeroth-order mean-field approximation to $\sum_{\mu_{\text{bond}=1}}^{z}\delta_{\text{bond}}$. Hence, $X_{\text{bond}}$ defines the bond correction term and reads
\begin{equation}
X_{\text{bond}}=\frac{N_l}{z}\left[ \sum_{\mu_{i}=1}^{z}\delta_{\text{bond}}(\mathbf{r}_{i},\mathbf{r}_{i+1}+\mathbf{a}_{\mu_{i}})-\frac{N_l}{z} \right].
\end{equation}
Processing the factors involving the nearest neighbor van der Waals interactions and chain semiflexibility in a similar fashion yields the partition function $W$ in the desired form of multiple cluster expansions,
\begin{eqnarray}
W=&&\sum_{\{\mathbf{r}\}}{\textstyle{'}}\frac{W^{mf}(N_l-mM)!}{N_{l}!}
\left[ \prod_{\text{bond}}\left( 1+X_{\text{bond}} \right) \right.\nonumber\\
&&
\times \prod_{\text{pair}} \left( 1+G_{\text{pair}}X_{\text{pair}} \right) \nonumber\\
&&
\times \left. \prod_{\text{bend}}\left( 1+K_{\text{bend}}Y_{\text{bend}} \right) \right],
\end{eqnarray}
where the interaction correction term $X_{\text{pair}}$ and its associated factor $G_{\text{pair}}$, and the bending correction term $Y_{\text{bend}}$ and its associated factor $K_{\text{bend}}$, appear as
\begin{equation}
X_{\text{pair}}=\frac{N_l}{z}\left[ \sum_{\mu_{i}=1}^{z}\delta_{\text{pair}}(\mathbf{r}_{i},\mathbf{r}_{j}+\mathbf{a}_{\mu_{i}})-\frac{N_l}{z} \right],
\end{equation}
\begin{equation}
G_{\text{pair}}=G=\frac{zf_{\text{pair}}/N_l}{1+zf_{\text{pair}}/N_l},
\end{equation}
\begin{equation}
Y_{\text{bend}}=z\delta_{\text{bend}}(\mu_{i},\mu_{i+1})-1,
\end{equation}
\begin{equation}
K_{\text{bend}}=K=\frac{1-E_{\text{bend}}}{(z-1)E_{\text{bend}}+1}.
\end{equation}

Expanding the product in Eq. (8) leads naturally to the cluster expansion for the partition function $W$ that bears some similarity to the Mayer cluster expansion in the theory of nonideal gases. Then, the partition function can be finally expressed as a zeroth-order mean-field term and a cluster expansion
\begin{eqnarray}
W=&&W^{mf}\left[ 1+\frac{(N_l-mM)!}{N_{l}!}\sum_{\{\mathbf{r}\}}{\textstyle{'}} \left(\sum_{i}t_{i}
\right. \right. \nonumber\\
&&
\left. \left. +\sum_{i,j}t_{i}t_{j}+\sum_{i,j,k}t_{i}t_{j}t_{k}+\cdots\right) \right],
\end{eqnarray}
where $t_{\alpha} (\alpha=i,j,k,...)$ is either $X_{\text{bond}}$, $G_{\text{pair}}X_{\text{pair}}$, or $K_{\text{bend}}Y_{\text{bend}}$. The cluster expansion for the bending constraints differs from the cluster expansions generated by the other two constraints because a factor of $Y_{\text{bend}}$ only contributes to the partition function when multiplied by the bond correlation corrections $X_{\text{bond}}$ for the two bonds affected by the factor $Y_{\text{bend}}$. 

\subsection{Diagrammatic representation}

The systematic calculation of contributions from the cluster corrections in Eq. (13) can be transformed into explicit expressions for evaluating contributions by representing each $t_{\alpha}$ factor diagrammatically~\cite{ACP_103_335, JCP_87_7272}. Consider first diagrams where the factors of $t_{\alpha}$ only contain factors of $X_{\text{bond}}$ or $G_{\text{pair}}X_{\text{pair}}$. Each factor of $X_{\text{bond}}$ represents a correlating bond and is depicted by a solid straight line connecting the bonded united atom groups, each represented by circles. Thus, diagrams containing only solid straight lines and circles are called bond diagrams. When the bond diagrams contain two portions separated by one or more intervening bonds on the same chain, a dotted straight line is used to indicate that the separated portions belong to the same chain. The same rule also applies to other types of diagrams with two separated bond portions. The present LCT retains all bond diagrams where the total number of correlating lines is less than or equal to four, and a full list of contributing bond diagrams can be found in Ref.~\cite{JCP_87_7272}. The pair diagrams depict each factor of the van der Waals interaction term $G_{\text{pair}}X_{\text{pair}}$ as a solid curved line connecting the pair of uncorrelated interacting united atom groups, which are represented by crosses and also called bare interacting united atom groups in the following since they differ from those united atom groups participating in the correlated bonds (called correlated interacting united atom groups). Likewise, the diagrams involving both $X_{\text{bond}}$ and $G_{\text{pair}}X_{\text{pair}}$ factors are termed bond/pair diagrams, where the correlated united atom group participating in the interactions is also designated as crosses instead of circles. Dudowicz and Freed~\cite{Mac_24_5076} provide expressions for all relevant pair and bond/pair diagrams containing up to a total of four bond and/or interaction lines. The bond and bond/pair diagrams provide the essential building blocks for constructing the remaining diagrams including the bending energy factor $K_{\text{bend}}Y_{\text{bend}}$, which are generated by adding dashed curved lines, connecting pairs of consecutive bond correlation lines in the relevant bond and bond/pair diagrams, as described in detail by Foreman and Freed~\cite{ACP_103_335}. 

The diagrammatic representation for the partition function then contains diagrams with all possible combinations of elements, including solid straight bond correlation lines, solid curved interaction lines, dashed curved bending constraint lines, correlated circles, interacting crosses, and dotted straight lines. In summary, the LCT calculations include contributions from the following five types of diagrams:
\begin{enumerate}
\item \textbf{Bond diagrams}, each of which consists of solid straight lines, circles, and possibly a dotted straight line;
\item \textbf{Pair diagrams}, each of which consists of solid curved lines and crosses;
\item \textbf{Bond/pair diagrams}, each of which consists of solid straight lines, solid curved lines, circles, crosses, and possibly a dotted straight line;
\item \textbf{Bond/bend diagrams}, each of which consists of solid straight lines, dashed curved lines, circles, and possibly a dotted straight line;
\item \textbf{Bond/pair/bend diagrams}, each of which consists of solid straight lines, solid curved lines, dashed curved lines, circles, crosses, and possibly a dotted straight line.
\end{enumerate}
In the following, the diagrams containing one or two solid curved interaction lines are also termed first-order or second-order interaction energy diagrams, respectively.

\begin{figure*}[tb]
 \centering
 \includegraphics[angle=0,width=0.8\textwidth]{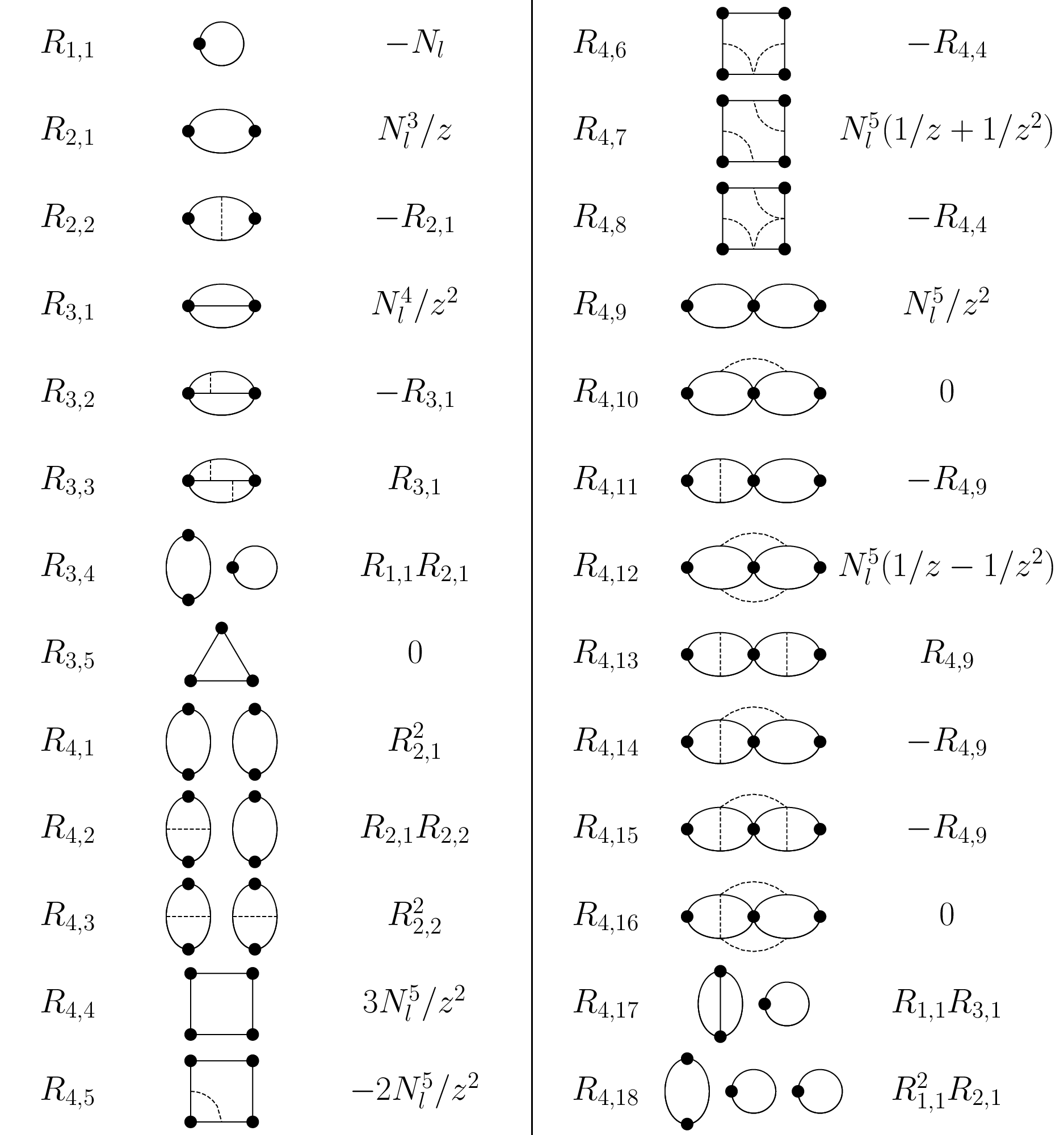}
 \caption{LCT contracted diagrams (middle) along with their labels (left) and expressions (right). The contracted diagrams depicted represent all possibilities with closed loops, and thus some vanishing contracted diagrams are also included for completeness. The solid lines emerge from the lines specifying the correlating bonds and the nearest neighbor interactions in the original diagram, while the dashed lines originate from the lines of the original diagram imposing the bending constraints.}
\end{figure*}

We now provide a general description for evaluating the diagrams of the LCT. The value of an individual diagram with $B$ solid straight lines, $l$ solid curved lines and $b$ dashed curved lines is the product of a monomer structure independent but lattice structure dependent connectivity factor $D(B, l, b)$ and a monomer structure dependent but lattice structure independent combinatorial factor $\gamma_{D}(B, l, b)$. Thus, the partition function given by Eq. (13) transforms into the more compact representation,
\begin{eqnarray}
W=W^{mf}\left[ 1+\sum_{B, l, b}\gamma_{D}(B, l, b)D(B, l, b) \right].
\end{eqnarray}

The connectivity factor $D(B, l, b)$ can be generally written for all types of diagrams as
\begin{eqnarray}
D(B, l, b)=\frac{d(B, l, b)}{\alpha}G^{l}K^{b},
\end{eqnarray}
where the denominator $\alpha$ only depends on the total number of lattice sites $N_l$ in the whole lattice and the number of vertices (i.e., united atom groups) $N_v$ in the diagram as
\begin{eqnarray}
\alpha=N_l(N_l-1)\cdots(N_l-N_v+1),
\end{eqnarray}
and the factors $G$ and $K$ are given by Eqs. (10) and (12). The factor $d(B, l, b)$ is dependent on the lattice structure but not on the chain architecture, i.e., monomer structure. The calculation of $d(B, l, b)$ is highly nontrivial due to the excluded volume constraints, and systematic methods have been developed in Ref.~\cite {JCP_87_7272} in terms of contracted diagrams, a set of new diagrams that are obtained by merging sets of two or more vertices for the given diagram to evaluate the corrections from the zeroth-order mean-field approximation that replaces these excluded volume constraints by the average of $\delta_{bond}$. Specifically, $d(B, l, b)$ for each diagram is evaluated in the form
\begin{eqnarray}
d(B, l, b)=\sum_{c}f_{B,c}R_{B,c},
\end{eqnarray}
where $c$ is a sequential counting index and the coefficient $f_{B,c}$ is the product of a contraction factor $\prod_{\lambda=1}^{N'_v}(-1)^{k_{\lambda}-1}(k_{\lambda}-1)$ and the number of ways of forming a contracted diagram with $N'_v$ vertices by merging the distinguishable vertices $k_1, k_2, ..., k_{N'_v}$ in the original uncontracted diagram. The expressions for the relevant contracted diagrams $R_{B,c}$ are provided in Fig. 2 with some corrections to the table in Ref.~\cite{ACP_103_335}. 

The combinatorial factor $\gamma_{D}(B, l, b)$ emerges in the form
 \begin{eqnarray}
\gamma_{D}(B, l, b)=s_D\gamma(B, l, b),
\end{eqnarray}
where $s_D$ is the symmetry number for the diagram and $\gamma(B, l, b)$ is the number of ways of selecting the set of correlating bonds as well as bare interacting united atom groups from all polymer chains in the system. The symmetry number $s_D$ depends on the nearest neighbor interaction energy factor, and, by definition, $s_D$ always equals unity for $l=0$. The computation of $s_D$ for $l>0$ proceeds by removing the interaction lines in the diagram and assigning labels to all united atom groups. Then, the symmetry number $s_D$ equals the number of distinguishable ways of constructing the final diagram by inserting the interaction lines. The computation of $s_D$ for diagrams with $l>0$ is detailed in Ref.~\cite{Mac_24_5076}.

The definition of $\gamma(B, l, b)$ immediately implies that the expression for the LCT partition function contains a set of geometrical counting indices, depending on the structure of the correlating bonds in the diagrams. Diagrams without bending constraints include contributions from all possible combinations of bonds that produce the cluster architecture depicted in the diagram. However, the treatment of the bending constraints requires further specifications for the counting indices because the bending constraints limit the number of combinations by constraining the affected pair of consecutive bonds to lie on the same subchain. Thus, the total number of sequential two-bond runs $N_2$ in a single chain must be the sum of the number of the bond pairs lying along the identical subchain $N_{2i}$ and the number of the bond pairs that reside in two different subchains $N_{2d}$, 
 \begin{eqnarray}  
N_2=N_{2i}+N_{2d}.
\end{eqnarray}
A single bending energy factor appears only in conjunction with $N_{2i}$. Details of the derivation of similar relations for the necessary counting indices can be found in Refs.~\cite{ACP_103_335, JCP_119_5730}.

\subsection{LCT free energy}

The LCT Helmholtz free energy $F$ follows from the partition function by applying the standard thermodynamic definition to Eq. (14)
 \begin{eqnarray}
\beta F=-\ln W=-\ln W^{mf}-\ln(1+\sigma),
\end{eqnarray}
where $\sigma=\sum_{B, l, b}\gamma_{D}(B, l, b)D(B, l, b)$. The last term of Eq. (20) can be expanded in a Taylor series
\begin{eqnarray}
\ln(1+\sigma)=\sigma-\frac{1}{2}\sigma^2+\frac{1}{3}\sigma^3-\cdots.
\end{eqnarray}
Recognizing that the Helmholtz free energy $F$ is extensive, all resulting terms in $F$ must be extensive and scale as the first power of $N_l$. Individual terms contributing to $\sigma$ might be hyperextensive, i.e., proportional to $N_l^{p}$ with $p>1$. Rearranging the series on the right-hand side of Eq. (21) into cumulants, each of which is a sum of several original diagrams, identically eliminates hyperextensive contributions~\cite{JCP_87_7272}. The diagrams with separated bond portions must also produce hyperextensive terms, whose eliminations arise from cancellation with the diagrams lacking separated bond portions.

The current LCT retains all bond diagrams through the second order in $1/z$. In addition, the Mayer $f$ functions are treated using a high temperature expansion
\begin{eqnarray}
f_{\text{pair}}=(\beta\epsilon_{ij})+\frac{1}{2!}(\beta\epsilon_{ij})^2+\frac{1}{3!}(\beta\epsilon_{ij})^3+\cdots,
\end{eqnarray}
where the first-order term are retained for all interaction energy diagrams, and the second-order term are retained only for several first-order interaction energy diagrams~\cite{Mac_24_5076}. The cluster expansion for the free energy is obtained treating $\epsilon_{ij}$ formally as order $1/z$.

Rearranging the expression for the LCT free energy $f=F/N_l$ in powers of the polymer volume fraction $\phi$ converts the expression into the general form,
\begin{equation}
\beta f=\beta f^{mf}-\sum_{i=1}^6C_i\phi^i.
\end{equation}
The term $\beta f^{mf}$ represents the zeroth-order mean-field contribution, which for a one-bending energy model reads
\begin{eqnarray}
\beta f^{mf}=&&\frac{\phi}{M}\ln(\frac{2\phi}{z^{L}M})+\phi(1-\frac{1}{M})+ (1-\phi)\ln(1-\phi)\nonumber\\
&&
-\phi \frac{N_{2i}}{M}\ln(z_b),
\end{eqnarray}
where $z_b=(z_p-1)\exp(-\beta E_b)+1$ with $z_p=z/2$. The choice of $z_p=z/2$ assumes the existence of one \textit{trans} and $z/2-1$ \textit{gauche} conformations for a pair of successive bonds. The coefficient $C_i$ in Eq. (23) is obtained by combining contributions from all relevant diagrams. Applying the above approximations to the model melt with monomer averaged interactions where all pairs of united atom groups have a common averaged van der Waals interaction energy $\epsilon$, $C_i$ appears in the compact form,
\begin{equation}
C_i=C_{i, 0}+C_{i, \epsilon}(\beta\epsilon)+C_{i, \epsilon^2}(\beta\epsilon)^2,
\end{equation}
where $C_{i, 0}$, $C_{i, \epsilon}$, and $C_{i, \epsilon^2}$ are obtained by combining terms independent of $\epsilon$, terms dependent on $\epsilon$, and terms dependent on $\epsilon^2$, respectively. Explicit expressions for all the coefficients $C_1,...,C_6$ in Appendix A contain corrections to the formulas in earlier works~\cite{ACP_103_335}.  Also note that the results given in Appendix A are specialized to the one-bending energy model, where a common value for the bending energy has been assumed. It is straightforward to generalize the results in Appendix A to models of melts with different bending energies for the backbone and side chains by using the relations provided in Appendix B of Ref.~\cite{JCP_124_064901}. 

\section{Extension of the lattice cluster theory to model polymer melts with specific interactions}

In this section, we describe in detail how to extend the LCT to models of polymer melts where the side chains' end segments are permitted to have different nearest neighbor van der Waals interaction energies with other united atom groups. In particular, we discuss the new features arising from the model melts with specific interactions. This section also provides typical examples for the evaluation of the diagrams in the LCT.

\subsection{Generalized counting indices and interaction energy diagrams}

\begin{figure}[tb]
 \centering
 \includegraphics[angle=0,width=0.35\textwidth]{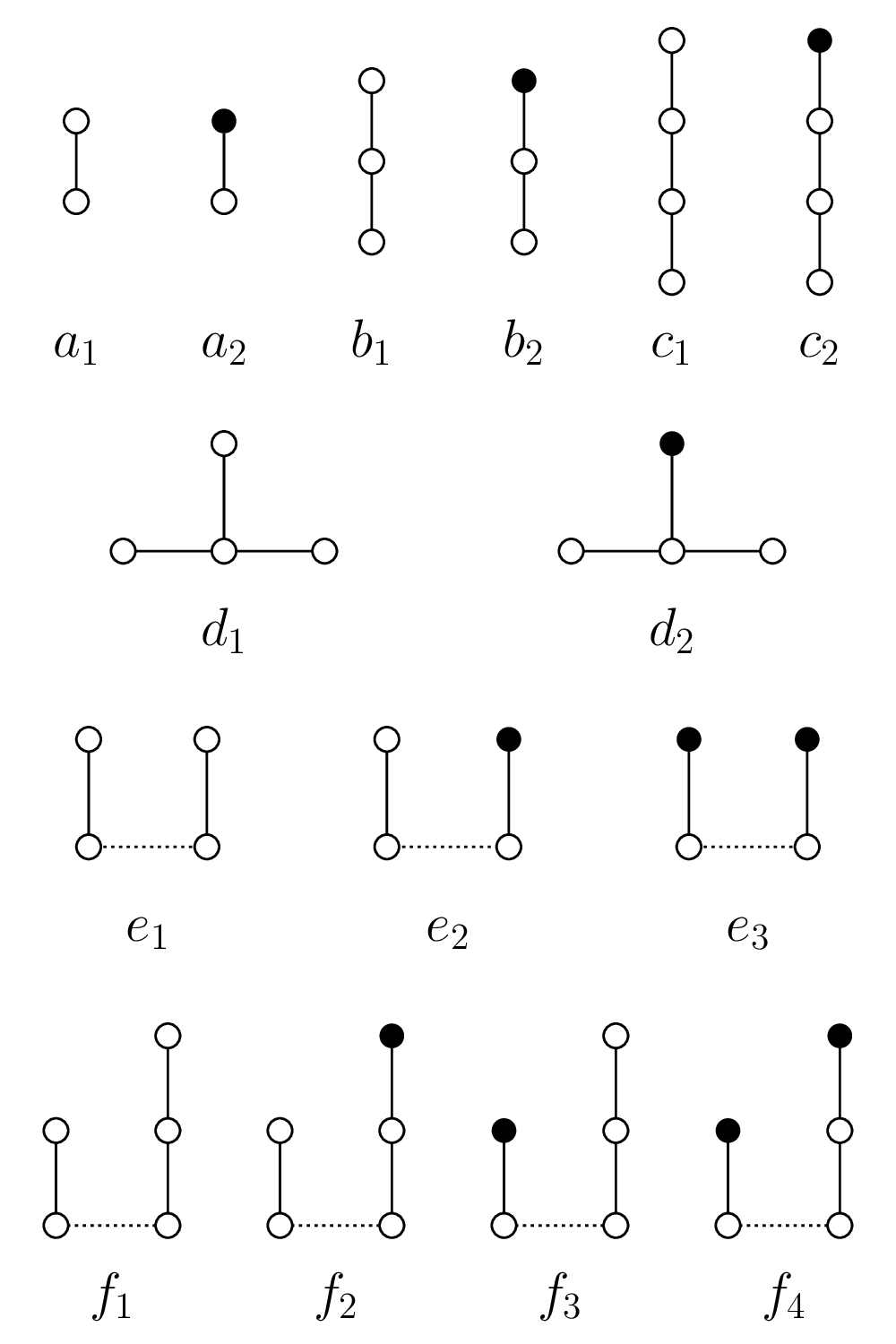}
 \caption{All configurations of bonds that constitute the bond/pair and bond/pair/bend diagrams for the model of melts with specific interactions. Open and solid circles correspond to normal united atom groups (i.e., n-groups) and to the end segments of the side chains (i.e., e-groups), respectively. Solid straight lines denote the bonds between pairs of bonded untied atom groups, while dotted straight lines in $e_1-e_3$ and $f_1-f_4$ indicate the presence of one or more intervening bonds between the depicted bonds, as explained in Subsection III B.}
\end{figure}

\begin{figure}[tb]
 \centering
 \includegraphics[angle=0,width=0.48\textwidth]{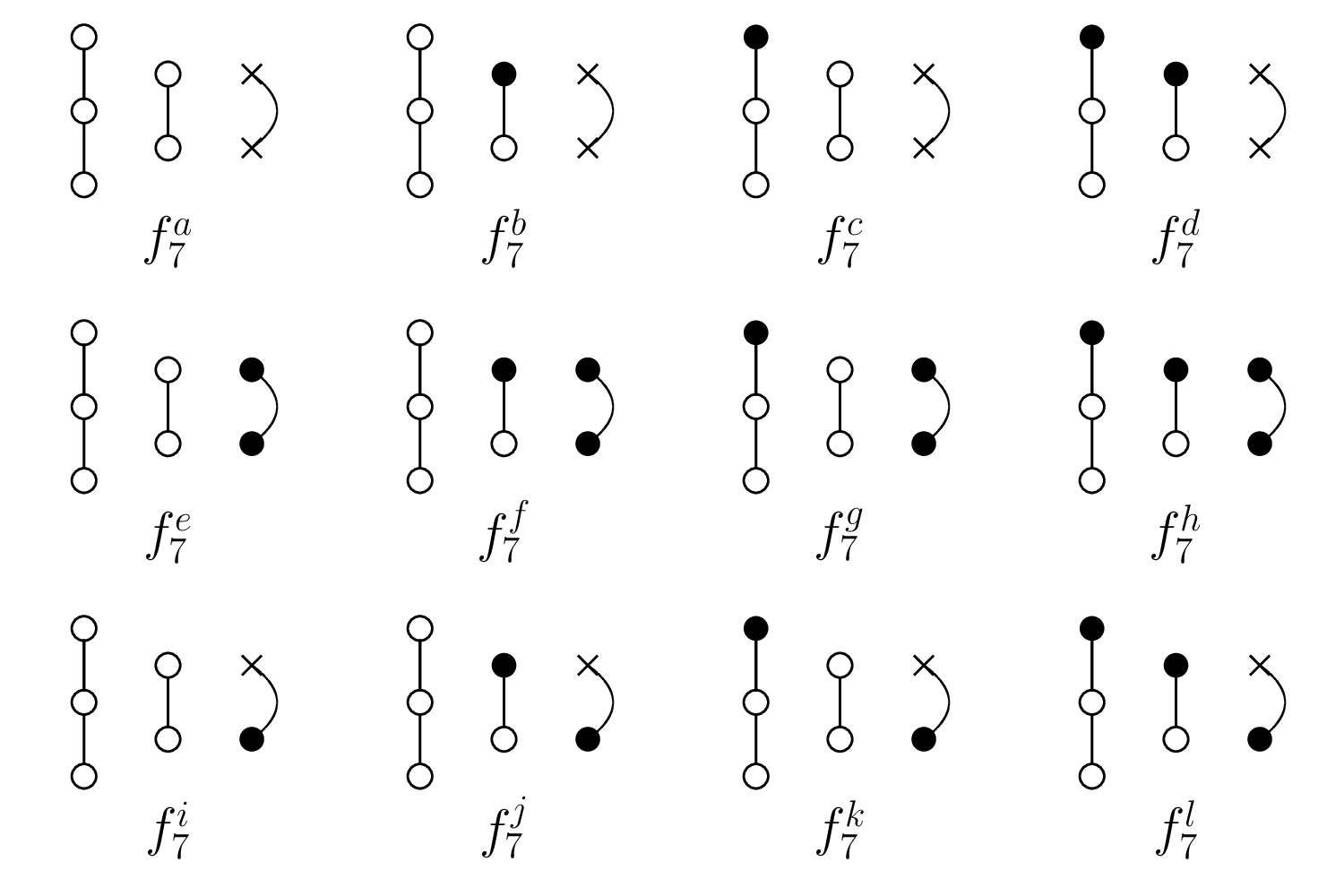}
 \caption{All possible bond/pair cumulant diagrams for the model of melts with specific interactions. All these diagrams may be derived from the first-order interaction energy diagram $f_7$ for models of melts with monomer averaged interactions by enumerating all possible ways in which united atom groups in the diagram may be specified as n-groups or e-groups.}
\end{figure}

Having provided the general formulation for deriving the LCT free energy in Section III, it is now straightforward to extend the theory to models of polymer melts with specific interactions, where the side chains' end segments (i.e., e-groups) differ from the other united atom groups (i.e., n-groups) in the strength of nearest neighbor interactions. Apparently, such a model does not require any modifications for the treatment of all constraints as illustrated in Subsection III A. The partition function can still be represented diagrammatically. The bond and the derived bond/bend diagrams for the melts with specific interactions are exactly the same as those in the melts with monomer averaged interactions because the interaction energy factors are absent in such diagrams. The diagrams involving the interaction energy factors, however, require systematic extensions.

We begin by describing the extensions necessary for constructing the bond/pair diagrams, i.e., diagrams involving both the bond correction factors $X_{\text{bond}}$ and the interaction energy factors $G_{\text{pair}}X_{\text{pair}}$. The six bond configurations, depicted by $a_1$, $b_1$, $c_1$, $d_1$, $e_1$ and $f_1$ in Fig. 3, are the basic elements of the bond/pair diagrams for the model melts with monomer averaged interactions, and each of them generates two or more distinguishable configurations in order to treat the model melts with specific interactions considered in the present work. For example, the two-bond configuration includes two specific cases for the model melts with specific interactions, namely, one configuration with all n-groups ($b_1$ in Fig. 3) and another with one e-group residing at either end ($b_2$ in Fig. 3). Thus, the counting index $N_2$ for the model melts with monomer averaged interactions is decomposed into two counting indices $N_{2n}$ and $N_{2e}$, which denote the numbers of two-bond runs in a single chain with all n-groups and of two-bond runs containing one e-group at either end, respectively. Figure 3 describes the decomposition of all six basic bond architectures into configurations necessary for the model melts with specific interactions considered in the present work. Accordingly, the decomposition of the counting indices yields the following relations,
\begin{subequations}
\begin{equation}
N_1=N_{1n}+N_{1e},
\end{equation}
\begin{equation}
N_2=N_{2n}+N_{2e},
\end{equation}
\begin{equation}
N_3=N_{3n}+N_{3e},
\end{equation}
\begin{equation}
N_{\bot}=N_{\bot n}+N_{\bot e},
\end{equation}
\begin{equation}
N_{1,1}=N_{1n,1n}+N_{1n,1e}+N_{1e,1e},
\end{equation}
\begin{equation}
N_{1,2}=N_{1n,2n}+N_{1n,2e}+N_{1e,2n}+N_{1e,2e}.
\end{equation}
\end{subequations}
Expressing the counting indices $N_{x,y}$ for diagrams with two disconnected portions $x$ and $y$ in terms of several individual counting indices~\cite{PRA_45_7111} simplifies the final expression for the free energy. Appendix B provides the expressions for $N_{1,1}$ and $N_{1,2}$ along with their decomposed forms for the model melts with specific interactions considered in the present work.

Likewise, each of bare interacting united atom groups in the bond/pair diagrams can be either a n-group or an e-group. Extending the interaction energy diagrams to models of melts with specific interactions then proceeds by counting all possible ways of combining the pieces comprising each individual diagram, a process illustrated for the first-order interaction energy diagram $f_7$. (A full list of the first-order interaction energy diagrams appears in Fig. 3 of Ref.~\cite{Mac_24_5076}). Diagram $f_7$ contains three pieces: a one-bond configuration, a two-bond configuration and two bare interacting united atom groups. All possibilities of each piece and their combinations convert diagram $f_7$ into twelve different generalized diagrams, as shown in Fig. 4. The bare interacting e-groups are no longer designated by crosses in order to distinguish them from the interacting n-groups when they participate in the nearest neighbor interactions. Diagrams $f_7^a-f_7^d$, $f_7^e-f_7^h$ and $f_7^i-f_7^l$ provide contributions for the interaction terms $\epsilon_{11}$, $\epsilon_{22}$ and $\epsilon_{12}$, respectively. The generalization of other bond/pair diagrams follows similar procedures. 

Once all the bond/pair diagrams are generalized, the bending energy factor can be included using the same method described in Subsection III B to generate all the generalized bond/pair/bend diagrams. Meanwhile, the counting indices in Eq. (26)  must be further extended by specifying whether or not the successive bonds lie along the identical subchain. Appendix B provides the necessary relations for the counting indices in order to treat these additional bending constraints.

\subsection{Illustration of the diagram evaluation}

\begin{figure}[tb]
 \centering
 \includegraphics[angle=0,width=0.4\textwidth]{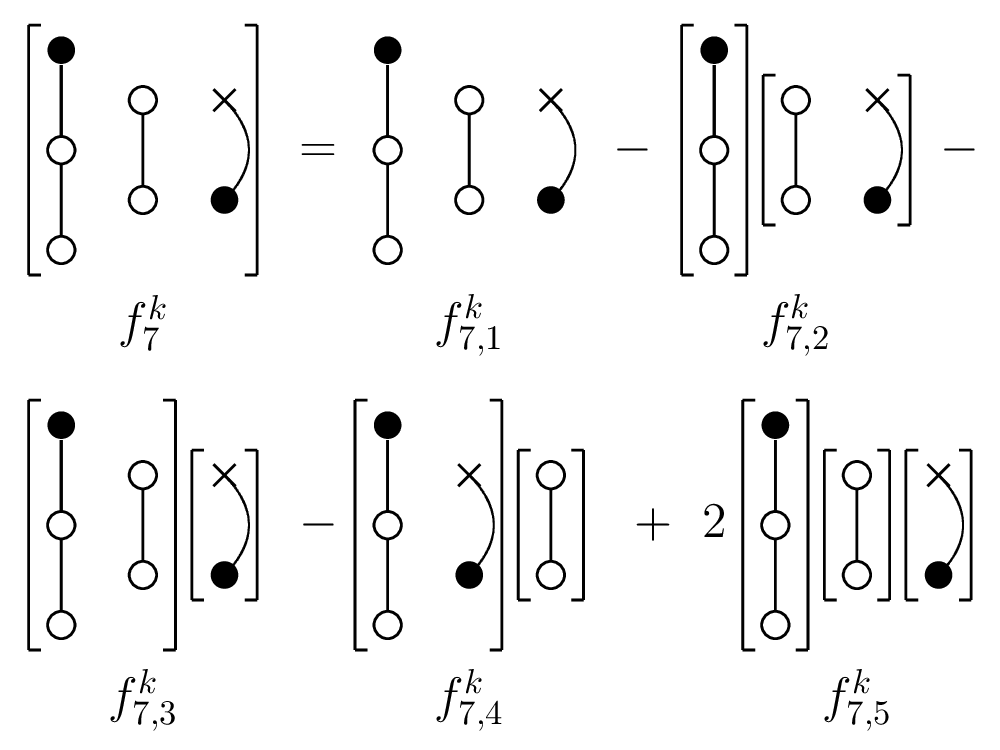}
 \caption{Illustration of the evaluation of the bond/pair diagram $f_7^k$ in Fig. 4. The cumulant diagram $f_7^k$ is composed of five individual diagrams $f_{7,1}^k$, $f_{7,2}^k$, $f_{7,3}^k$, $f_{7,4}^k$ and $f_{7,5}^k$, whose combination yields the final result for $f_7^k$.}
\end{figure}

\begin{figure}[tb]
 \centering
 \includegraphics[angle=0,width=0.4\textwidth]{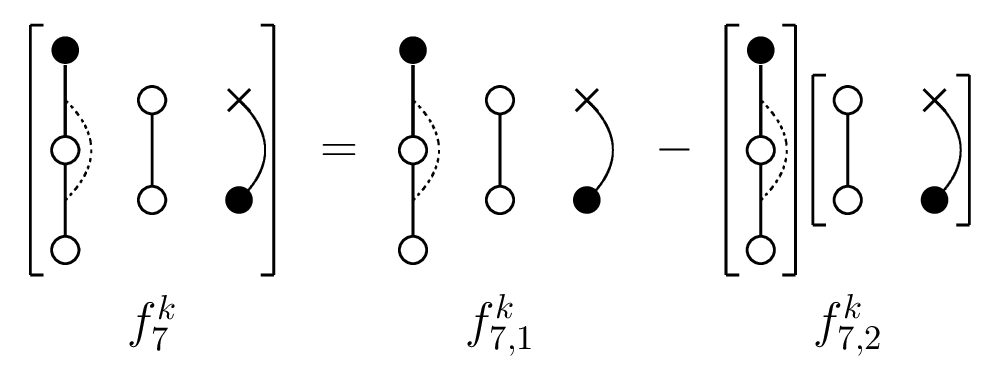}
 \caption{Illustration of the evaluation of the bond/pair/bend diagram derived from $f_7^k$ in Fig. 4. Only the first two individual diagrams survive when including the bending energy factor.}
\end{figure}

\begin{table*}[tb]
\caption{Expressions of the terms that are necessary for the evaluation of each individual diagram and hence the cumulant bond/pair diagram $f_7^k$ in Fig. 5.}
\begin{ruledtabular}
\begin{tabular}{llll}
\textrm{Diagram}&
\textrm{$\alpha$}&
\textrm{$d$}&
\textrm{$\gamma_D$}\\
\colrule
$f_{7,1}^{k}$ & $\prod_{i=0}^{6}(N_l-i)$ & $-2R_{4,1}-8R_{4,2}+48R_{4,3}$ & $N_{1n}N_{2e}m(m-1)(f_nmM-4)(f_emM-1)$\\
~ & ~ & $+8R_{4,4}-R_{4,5}$ & ~\\
$f_{7,2}^{k}$ & $\prod_{i=0}^{2}(N_l-i)\prod_{i=0}^{3}(N_l-i)$ & $-2R_{4,1}-R_{4,5}$ & $N_{1n}N_{2e}m^2(f_nmM-2)(f_emM)$\\
$f_{7,3}^{k}$ & $\prod_{i=0}^{4}(N_l-i)\prod_{i=0}^{1}(N_l-i)$ & $4R_{4,4}-R_{4,5}$ & $N_{1n}N_{2e}m(m-1)(f_nmM-2)(f_emM)$\\
$f_{7,4}^{k}$ & $\prod_{i=0}^{4}(N_l-i)\prod_{i=0}^{1}(N_l-i)$ & $4R_{4,4}-R_{4,5}$ & $N_{1n}N_{2e}m^2(f_nmM-2)(f_emM-1)$\\
$f_{7,5}^{k}$ & $\prod_{i=0}^{2}(N_l-i)[\prod_{i=0}^{1}(N_l-i)]^2$ & $-R_{4,5}$ & $N_{1n}N_{2e}m^2(f_nmM)(f_emM)$\\
\end{tabular}
\end{ruledtabular}
\end{table*}

\begin{table*}[tb]
\caption{Expressions of the terms that are necessary for the evaluation of each individual diagram and hence the cumulant bond/pair/bend diagram $f_7^k$ in Fig. 6.}
\begin{ruledtabular}
\begin{tabular}{llll}
\textrm{Diagram}&
\textrm{$\alpha$}&
\textrm{$d$}&
\textrm{$\gamma_D$}\\
\colrule
$f_{7,1}^{k}$ & $\prod_{i=0}^{6}(N_l-i)$ & $-2R_{4,2}-8R_{4,5}+16R_{4,10}$ & $N_{1n}N_{2ei}m(m-1)(f_nmM-4)(f_emM-1)$\\
~ & ~ & $+32R_{4,11}$ & ~\\
$f_{7,2}^{k}$ & $\prod_{i=0}^{2}(N_l-i)\prod_{i=0}^{3}(N_l-i)$ & $-2R_{4,2}$ & $N_{1n}N_{2ei}m^2(f_nmM-2)(f_emM)$\\
\end{tabular}
\end{ruledtabular}
\end{table*}

The evaluation of generalized diagrams follows exactly the same method described in Subsection III B. The computation of each individual diagram with $B$ bond correlation lines, $l$ interaction lines and $b$ bending constraint lines requires knowledge of the connectivity factor $D(B, l, b)$ and the combinatorial factor $\gamma_{D}(B, l, b)$. These two factors, $D(B, l, b)$ and $\gamma_{D}(B, l, b)$, in turn, can be calculated through Eqs. (15-18). The bond/pair diagram $f_7^k$ in Fig. 4 is selected to illustrate the evaluation process and further clarify the individual contributions to $D(B, l, b)$ and $\gamma_{D}(B, l, b)$.

The cumulant diagram $f_7^k$ is a combination of five individual diagrams, as depicted in Fig. 5. The topology of each individual diagram can only be constructed in one possible manner by removing the solid curved interaction line and labeling all united atom groups. Therefore, the symmetry number is $s_D=1$ for all the five individual diagrams. Table I provides a summary for each term of the other factors in $D(B, l, b)$ and $\gamma_{D}(B, l, b)$. The diagram $f_{7,1}^k$ is used as an example to explain the meaning of each term. The diagram $f_{7,1}^k$ contains seven united atom groups, and thus the factor $\alpha$ simply reads $\alpha=\prod_{i=1}^{6}(N_l-i)$. The factor $d=-2R_{4,1}-8R_{4,2}+48R_{4,3}+8R_{4,4}-R_{4,5}$ with the values of $R_{B,c}$ given in Fig. 2 is obtained from the contracted diagrams generated when the interaction line is replaced by a correlating bond. The factor $\gamma=N_{1n}N_{2e}m(m-1)(f_nmM-4)(f_emM-1)$ is the number of ways of selecting the combination of bonds and bare interacting united atom groups from all polymer chains in the system.

Combining the results from all five individual diagrams according to the rule given in Fig. 5 leads to the contribution to the free energy from the cumulant bond/pair diagram $f_7^k$ as
\begin{eqnarray}
f_7^k(3,1,0)\rightarrow&&\frac{N_l}{z}\epsilon_{12}\left[\phi^3(2f_n+4f_e+2f_nf_eM)\right]\nonumber\\
&&
\times \frac{N_{1n}N_{2e}}{M^2},
\end{eqnarray}
where the numbers $(3,1,0)$ in the bracketed term correspond to the values of $B$, $l$ and $b$, and serve to distinguish different types of diagrams. The same notation is also used in the following for the calculation of a bond/pair/bend diagram. The hyperextensive term (i.e., the term with $2f_nf_eM$) cancels identically with that arising from the bond/pair diagram containing the counting index $N_{1n,2e}$.

The evaluation of diagrams with bending energy factors proceeds in a similar fashion. Since the bending energy factors introduce additional constraints, the cumulant diagram $f_7^k$ with the bending energy factor is now a combination of only two individual diagrams, as shown in Fig. 6. The explicit forms of $\alpha$, $d$ and $\gamma_D$ are summarized in Table II for individual diagrams, and the final contribution to the free energy from the cumulant bond/pair/bend diagram $f_7^k$ is
\begin{eqnarray}
f_7^k(3,1,1)\rightarrow&&\frac{N_l}{z}\epsilon_{12}K \left[\phi^3(-2f_n-4f_e-2f_nf_eM) \right.\nonumber\\
&&
\left.+\phi^4(8f_nf_e) \right]\frac{N_{1n}N_{2ei}}{M^2}.
\end{eqnarray}
A bond/pair/bend diagram containing the counting factor $N_{1n,2ei}$ cancels the hyperextensive term in Eq. (28). This typical example explains how inclusion of the bending energy factor introduces a new parameter $K$ and also additional powers in $\phi$.

\subsection{LCT free energy for the model polymer melts with specific interactions}

The LCT Helmholtz free energy for the model polymer melts with specific interactions still possesses the general form of Eq. (23), and the zeroth-order mean-field term remains unchanged. The coefficient $C_i$ becomes more lengthy because there are now three nearest neighbor van der Waals energies. Again, $C_i$ can be generally written as
\begin{eqnarray}
C_i=&&C_{i, 0}+C_{i, \epsilon_{11}}(\beta\epsilon_{11})+C_{i, \epsilon_{22}}(\beta\epsilon_{22})+C_{i, \epsilon_{12}}(\beta\epsilon_{12})\nonumber\\
&&
+C_{i, \epsilon_{11}^2}(\beta\epsilon_{11})^2+C_{i, \epsilon_{22}^2}(\beta\epsilon_{22})^2+C_{i, \epsilon_{12}^2}(\beta\epsilon_{12})^2\nonumber\\
&&
+C_{i, \epsilon_{11}\epsilon_{12}}(\beta\epsilon_{11})(\beta\epsilon_{12})+C_{i, \epsilon_{11}\epsilon_{22}}(\beta\epsilon_{11})(\beta\epsilon_{22})\nonumber\\
&&
+C_{i, \epsilon_{12}\epsilon_{22}}(\beta\epsilon_{12})(\beta\epsilon_{22}),
\end{eqnarray}
where the terms $C_{i, 0}$ are exactly identical to those for the model melts with monomer average interactions (see Appendix A for the explicit expressions of $C_{i, 0}$). Since the main aim of the present paper is to provide a general description for treating a class of models of polymer melts with specific interactions (e.g., similar methods can be applied to block copolymers),  the explicit expressions for the coefficients in Eq. (29), specializing to the model considered in the present work, are provided in Appendix C.

Since the free energy for the model of melts with specific interactions reduces to that for the model of melts with monomer averaged interactions when all three interaction energies are identical, the following relations must hold
\begin{eqnarray}
C_{i, \epsilon}=C_{i, \epsilon_{11}}+C_{i, \epsilon_{22}}+C_{i, \epsilon_{12}},
\end{eqnarray}
\begin{eqnarray}
C_{i, \epsilon^2}=&&C_{i, \epsilon_{11}^2}+C_{i, \epsilon_{22}^2}+C_{i, \epsilon_{12}^2}
+C_{i, \epsilon_{11}\epsilon_{12}}\nonumber\\
&&
+C_{i, \epsilon_{11}\epsilon_{22}}+C_{i, \epsilon_{12}\epsilon_{22}},
\end{eqnarray}
where $C_{i, \epsilon}$ and $C_{i, \epsilon^2}$ are terms in Eq. (25). Therefore, the models of melts with monomer averaged interactions indeed represent for a special case of the model with specific interactions.

\section{Illustration of the influence of specific interactions on the thermodynamic properties of polymer melts}

\begin{figure}[tb]
 \centering
 \includegraphics[angle=0,width=0.45\textwidth]{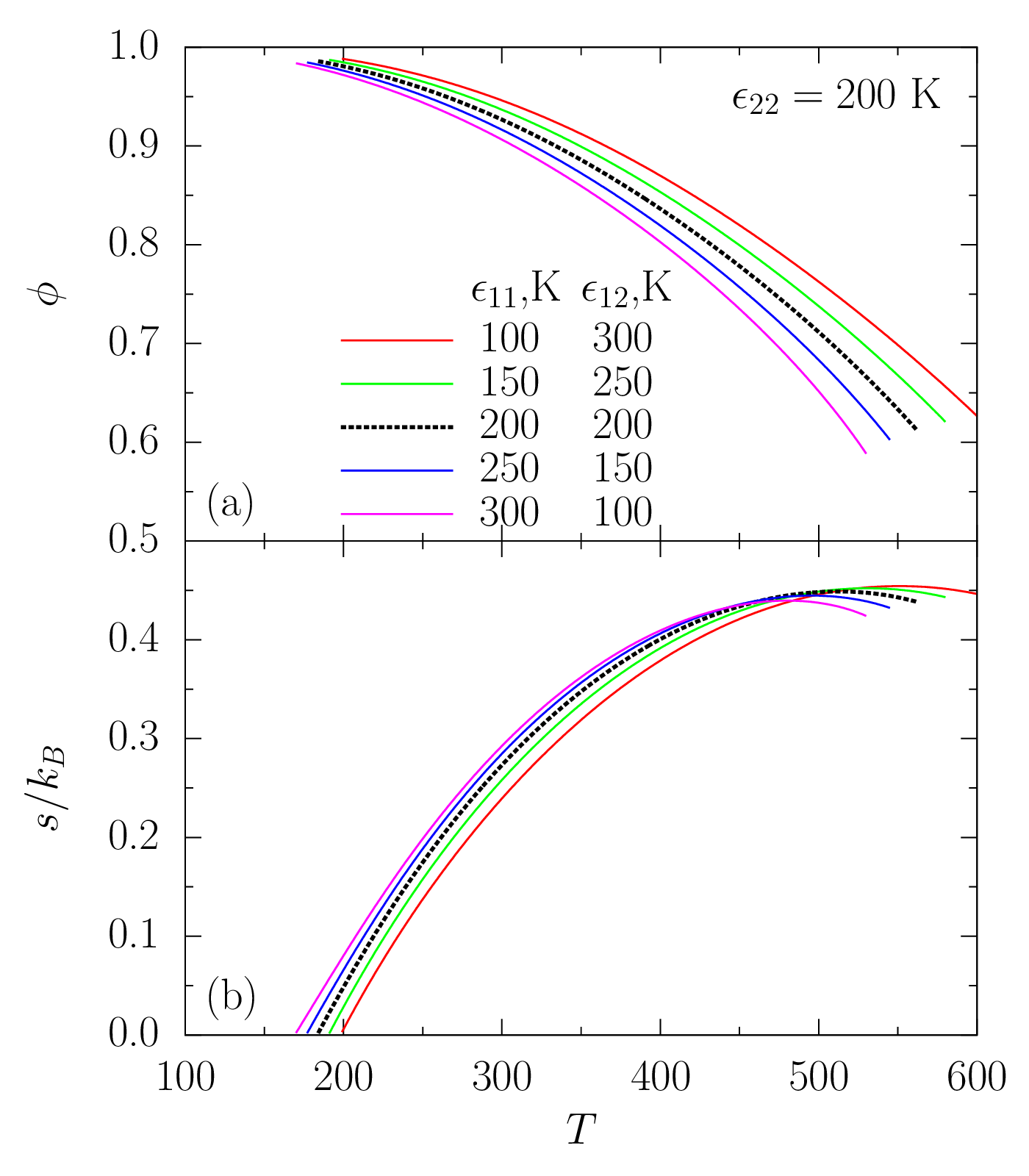}
 \caption{Illustration of the influence of specific interactions on the thermodynamic properties of polymer melts. The chains have the structure of poly(propylene) (PP) chains with polymerization index $N_c=8000$ and bending energy $E_b=400$ K. The figure displays the temperature $T$ dependence of the polymer volume fraction $\phi$ and the entropy $s/k_B$ at a constant pressure of $P=1$ atm in (a) and (b), respectively. The energy parameter $\epsilon_{22}$ is fixed as $200$ K, and the other two energy parameters $\epsilon_{11}$ and $\epsilon_{12}$ are adjusted according to Eq. (32) to produce an average interaction energy of $\epsilon_{\text{av}}=200$ K. The dashed lines highlight the results corresponding to the model with a monomer averaged interaction energy.}
\end{figure}

Although the monomer averaged interaction model suffices in establishing many qualitative trends, this model clearly represents a simplification that limits assessing the ability of using the energetic heterogeneity, which is ubiquitous within real monomers, as a further tool for designing materials with desirable thermodynamic properties. Thus, the current extension of the LCT to polymer melts with specific interactions clearly provides a more powerful vehicle for understanding and thereby controlling the thermodynamic properties of real polymers. This section illustrates the degree to which the thermodynamic properties can be modified by specific interactions even when constraining the three energy parameters $\epsilon_{11}$, $\epsilon_{22}$ and $\epsilon_{12}$ to a common average value, where the averaged interaction energy $\epsilon_{\text{av}}$ is defined as
\begin{eqnarray}
\epsilon_{\text{av}}=f_n^2\epsilon_{11}+f_e^2\epsilon_{22}+2f_nf_e\epsilon_{12},
\end{eqnarray}
and thus limits the number of free interaction energy parameters from three to two.

The PP structure (where the numbers of united atom groups per side chain and subchains per chain are $N_s=1$ and $L=N_c+1$) is chosen for the calculations because it requires the minimal number of parameters in the LCT. The illustrative calculations consider the equation of state (EOS), i.e., the temperature $T$ dependence of the polymer volume fraction $\phi$ at constant pressure, and the entropy $s$. Analytic expressions for the pressure $P$ and the entropy density $s$ (i.e., the entropy per lattice site) simply follow from the LCT Helmholtz free energy as
\begin{eqnarray}
P=-\left.\frac{\partial F}{\partial V}\right|_{m,T}=-\left.\frac{1}{V_{\text{cell}}}\frac{\partial F}{\partial N_l}\right|_{m,T},
\end{eqnarray}
and
\begin{eqnarray}
s=-\left. \frac{\partial f}{\partial T}\right|_{\phi},
\end{eqnarray}
where $V$ is the volume of the system, and $V_{\text{cell}}=a_{\text{cell}}^3$ is the volume associated with a single lattice site. The parameters $z=6$, $a_{\text{cell}}=2.7$\AA{}, $N_c=8000$, and $E_b=400$ K remain fixed in the calculations, and $\epsilon_{22}$ is kept as $200$ K while adjusting the other two energy parameters $\epsilon_{11}$ and $\epsilon_{12}$ according to Eq. (32) to produce $\epsilon_{\text{av}}=200$ K. The calculations illustrate the degree to which specific interactions yield properties that differ from the models of melts with the monomer averaged interaction energy of $\epsilon_{11}=200$ K. Figure 7 displays the $T$ dependence of $\phi$ and $s/k_B$ at a constant pressure of $P=1$ atm, and the dashed lines highlight the results for the monomer averaged interaction model. The monomer averaged interaction model captures the general trends of both quantities, but, more importantly, thermodynamic properties, such as the EOS and entropy density, can be modified somewhat by varying the specific interactions even when the averaged interaction energy $\epsilon_{\text{av}}$ is fixed. A separate work will analyze in detail the greater control of polymer properties that becomes possible when the averaged interaction energy is variable. Since both EOS and entropy density are important quantities entering into the description of polymer glass formation (see Ref.~\cite{ACP_137_125} for discussions of both quantities and for explanations why the entropy density nonmonotonically depends on temperature), the results presented in Fig. 7 suggest that the specific interactions provide additional variables for tailoring the properties of glass-forming polymers. 

\section{Discussion}

We extend the LCT to evaluate the Helmholtz free energy of models of polymer melts where the monomers are endowed with the $n$-$\alpha$-olefin structure and where the end segments of the side chains are permitted to have distinct, specific van der Waals interaction energies. The same methods may generally be applied to models in which all or certain of the side groups have distinct interaction energies from the backbone, the chain has other architectures, etc., but the details of the calculation differ in each case. However, the free energy expression for each case applies for many variables, including molecular weights, interaction energies, etc. The extended LCT free energy for the present example is a function of three nearest neighbor van der Waals interaction energies and reduces to the result for the model melts with monomer averaged interactions where all interaction energies are identical. 

Our work is largely motivated by experimental and simulational results~\cite{Mac_26_6824, Mac_31_4581, JCP_122_134505, Mac_41_7232,  Mac_45_8430, PCCP_15_4604, Lan_29_12730, JCP_140_044901}, demonstrating that specific interactions profoundly affect glass formation in polymeric systems. For instance, recent experiments~\cite{Mac_45_8430} indicate that both the fragility and the glass transition temperature can largely be tuned by modifying the chemical structure of the backbone and side groups and/or by controlling the spatial positions of the side groups. These studies illustrate the important changes in properties that accompany variations in monomer structure and molecular interactions and thus emphasize the importance of molecular factors in assisting the rational design of polymer materials. Our illustrative calculations imply that specific interactions provide additional variables for tailoring the thermodynamic properties of polymer melts and hence hold potential for controlling the properties of glass-forming polymers. The present extension of the LCT will be used in a separate paper to provide a similar generalization of the generalized entropy theory of polymer glass formation~\cite{ACP_137_125} and thus an improved understanding of the implications of specific interactions in glass formation of real polymers.

\begin{acknowledgments}
This work is supported by  the U.S. Department of Energy, Office of Basic Energy Sciences, Division of Materials Sciences and Engineering under Award DE-SC0008631.
\end{acknowledgments}

\appendix

\section{Summary of the coefficients that appear in the LCT free energy for the model melts with monomer averaged interactions}

\begin{table*}[tb]
\caption{Expressions for the coefficients that appear in the LCT free energy for the model melts with monomer averaged interactions where a single van der Waals energy applies to all pairs of united atom groups.}
\begin{ruledtabular}
\begin{tabular}{llll}
\textrm{$i$}&
\textrm{$C_{i,0}$}&
\textrm{$C_{i,\epsilon}$}&
\textrm{$C_{i,\epsilon^2}$}\\
\colrule
$1$ & $-u_{t2}/z-[u_{\bot}+C(u_2)+C(u_{o1})-C(u_4)]/z^2$ & $u_1+u_{t3}/z$ & $0$\\
$2$ & $u_1^2/z+(u_{t2}^2+2u_1u_{t3})/z^2$ & $z/2-2u_1+(2u_1^2-4u_1u_{t2}-2u_{t3})/z$ & $z/4-u_1+u_1^2+u_{t2}$\\
$3$ & $(2u_1^3/3-4u_1^2u_{t2})/z^2$ & $u_1+(-4u_1^2+4u_1^3+8u_1u_{t2}+u_{t3})/z$ & $-z/2+7u_1/2-6u_1^2-3u_{t2}$\\
$4$ & $2u_1^4/z^2$ & $(2u_1^2-8u_1^3-4u_1u_{t2})/z$ & $z/4-4u_1+12u_1^2+3u_{t2}$\\
$5$ & $0$ & $4u_1^3/z$ & $3u_1/2-10u_1^2-u_{t2}$\\
$6$ & $0$ & $0$ & $3u_1^2$\\
\end{tabular}
\end{ruledtabular}
\end{table*}

This appendix provides the explicit expressions for the coefficients that appear in the LCT free energy for the model melts with monomer averaged interactions. The results are summarized in Table III. Note that Table III details simplifying relations for the counting indices associated with diagrams containing disconnected portions within the same chain. The relations are expressed for convenience using the variables $u_x=N_x/M$ with $N_x$ denoting the number of runs of a given architecture $x$ in a single chain. Also, we have employed the following convenient expressions,
\begin{equation}
u_{t2}=u_{2i}g+u_{2d},
\end{equation}
\begin{equation}
u_{t3}=u_{3ii}g^2+u_{3id}g+u_{3dd},
\end{equation}
where $g=1-K=z_p\exp(-\beta E_b)/z_b$ (with the definitions of $z_p$ and $z_b$ given in Subsection III C of the main text) is another form of the bending energy factor. The shorthand terms that appear in $C_{1, 0}$ are given in the following,
\begin{equation}
C(u_2)=u_{2i}g^2/2+u_{2d}/2,
\end{equation}
\begin{equation}
C(o1)=u_{o1ii}(1-g)^2,
\end{equation}
\begin{eqnarray}
C(u_4)=&&-u_{4iii}g^3+(2u_{4iii}-u_{4idi})g^2\nonumber\\
&&
-(3u_{4iii}+2u_{4iid}-u_{4idi}+u_{4idd}\nonumber\\
&&
+u_{4did})g+u_{4iii}+u_{4iid}-u_{4idi}\nonumber\\
&&
-u_{4ddd},
\end{eqnarray}
where $u_{o1ii}=N_{o1ii}/M$ and $N_{o1}$ denotes the number of runs of the architecture in a single chain, corresponding to the bond diagram $o1$ in Fig. 5.2 of Ref.~\cite{ACP_103_335}, and $ii$ indicates that the three sequential bonds lie along one identical subchain. Note that $C(o1)$ and $C(u_4)$ provide the corrections to the previous results in Ref.~\cite{ACP_103_335}. Also note that the results given in Table III are specific to the one-bending energy model, where a common value for the bending energy has been assumed. However, the bending energy may be different for the backbone and side chains. It is straightforward to generalize the results to such a two-bending model by using the relations provided in Appendix B of Ref.~\cite{JCP_124_064901}. 

\section{Extension of the counting indices}

The previous LCT~\cite{PRA_45_7111} demonstrates that the counting index $N_{x,y}$ with two disconnected portions $x$ and $y$ can be always expressed in terms of several individual counting indices. For example, the following relations hold for any chain architecture,
\begin{equation}
2!N_{1,1}=N_1^2-N_1-2N_2,
\end{equation}
\begin{equation}
N_{1,2}=N_1N_2-2N_2-2N_3-3N_{\bot}.
\end{equation}

The derivation of similar relations becomes more complicated for the model melts with specific interactions. The final relations are provided in this Appendix where the counting indices constituting $N_{1,1}$ and $N_{1,2}$ are further specialized as to whether or not a set of bonds contain an e-group, 
%\begin{subequations}
\begin{equation*}
2!N_{1n,1n}=N_{1n}^2-N_{1n}-2N_{2n},
\end{equation*}
\begin{equation*}
N_{1n,1e}=N_{1n}N_{1e}-N_{2e},
\end{equation*}
\begin{equation}
2!N_{1e,1e}=N_{1e}^2-N_{1e},
\end{equation}
%\end{subequations}

%\begin{subequations}
\begin{equation*}
N_{1n,2n}=N_{1n}N_{2n}-2N_{2n}-3N_{3n}-3N_{\bot n},
\end{equation*}
\begin{equation*}
N_{1n,2e}=N_{1n}N_{2e}-N_{2e}-N_{3e}-2N_{\bot e},
\end{equation*}
\begin{equation*}
N_{1e,2n}=N_{1e}N_{2n}-N_{3e}-N_{\bot e},
\end{equation*}
\begin{equation}
N_{1e,2e}=N_{1e}N_{2e}-N_{2e},
\end{equation}
%\end{subequations}
where the subscript $n$ or $e$ denotes the absence or presence of an e-group at the end of the corresponding bond configuration (see Fig. 3).

The treatment of the bending energy factors requires further decomposition of the counting indices as to whether pairs of successive bonds lie on the identical subchain (labeled with $i$) or different subchains (labeled with $d$). The model melts with monomer averaged interactions requires the following relations,
\begin{equation}
N_{2}=N_{2i}+N_{2d},
\end{equation}
\begin{equation}
N_{3}=N_{3ii}+N_{3id}+N_{3dd},
\end{equation}
\begin{equation}
N_{\bot}=N_{\bot i}+N_{\bot d},
\end{equation}
\begin{equation}
N_{1,2}=N_{1,2i}+N_{1,2d}.
\end{equation}

The above relations are now generalized to the model melts with specific interactions considered in the present work as
%\begin{subequations}
\begin{equation*}
N_{2n}=N_{2ni}+N_{2nd},
\end{equation*}
\begin{equation}
N_{2e}=N_{2ei}+N_{2ed},
\end{equation}
%\end{subequations}

%\begin{subequations}
\begin{equation*}
N_{3n}=N_{3nii}+N_{3nid}+N_{3ndd},
\end{equation*}
\begin{equation}
N_{3e}=N_{3eii}+N_{3eid1}+N_{3eid2}+N_{3edd},
\end{equation}
%\end{subequations}

%\begin{subequations}
\begin{equation*}
N_{\bot n}=N_{\bot ni}+N_{\bot nd},
\end{equation*}
\begin{equation}
N_{\bot e}=N_{\bot ei}+N_{\bot ed},
\end{equation}
%\end{subequations}

%\begin{subequations}
\begin{equation*}
N_{1n,2n}=N_{1n,2ni}+N_{1n,2nd},
\end{equation*}
\begin{equation*}
N_{1n,2e}=N_{1n,2ei}+N_{1n,2ed},
\end{equation*}
\begin{equation*}
N_{1e,2n}=N_{1e,2ni}+N_{1e,2nd},
\end{equation*}
\begin{equation}
N_{1e,2e}=N_{1e,2ei}+N_{1e,2ed},
\end{equation}
%\end{subequations}

%\begin{subequations}
\begin{eqnarray*}
N_{1n,2ni}=&&N_{1n}N_{2ni}-2N_{2ni}-2N_{3nii}\nonumber\\
&&
-N_{3nid}-N_{\bot ni},
\end{eqnarray*}
\begin{equation*}
N_{1n,2ei}=N_{1n}N_{2ei}-N_{2ei}-N_{3eii}-N_{3eid1},
\end{equation*}
\begin{equation*}
N_{1e,2ni}=N_{1e}N_{2ni}-N_{3eii}-N_{3eid2}-N_{\bot ei},
\end{equation*}
\begin{equation}
N_{1e,2ei}=N_{1e}N_{2ei}-N_{2ei}.
\end{equation}
%\end{subequations}

Special attention is required to extend the counting index $N_{3e}$ because two possibilities exist for $N_{3eid}$ due to the presence of an e-group, namely, $N_{3eid1}$ and $N_{3eid2}$. $N_{3eid1}$ is the number of three-bond runs in a single chain, where the two bonds connecting the e-group lie along the same subchain, while $N_{3eid2}$ denotes the number of three-bond runs in a single chain when the two bonds composed of all n-groups lie along the same subchain.

\section{Summary of the coefficients that appear in the LCT free energy for the model melts with specific interactions}

This appendix provides the explicit expressions for the coefficients that appear in the LCT free energy for the model melts with specific interactions. The terms $C_{i,0}$ in Eq. (29) are identical to those given in Table III of the Appendix A, and the remaining terms in Eq. (29) are summarized in the following,
%\begin{subequations}
\begin{eqnarray*}
C_{1, \epsilon_{11}}=u_{1n}+\frac{1}{z}(u_{t3n}+u_{x3e}),
\end{eqnarray*}
\begin{eqnarray*}
C_{1, \epsilon_{22}}=0,
\end{eqnarray*}
\begin{eqnarray*}
C_{1, \epsilon_{12}}=u_{1e}+\frac{1}{z}(u_{t3e}-u_{x3e}),
\end{eqnarray*}
\begin{eqnarray*}
C_{1, \epsilon_{11}^2}=0, C_{1, \epsilon_{22}^2}=0, C_{1, \epsilon_{12}^2}=0,
\end{eqnarray*}
\begin{eqnarray}
C_{1, \epsilon_{11}\epsilon_{12}}=0, C_{1, \epsilon_{11}\epsilon_{22}}=0, C_{1, \epsilon_{12}\epsilon_{22}}=0,
\end{eqnarray}
%\end{subequations}

%\begin{subequations}
\begin{eqnarray*}
C_{2, \epsilon_{11}}=&&f_n^2z/2-f_n(u_1+u_{1n})+\frac{1}{z}[f_n(-u_{t3}\nonumber\\
&&
-u_{t3n}-u_{x3e})+2u_1u_{1n}+u_{1e}^2/2\nonumber\\
&&
-(u_1+u_{1n})(u_{t2}+u_{t2n})],
\end{eqnarray*}
\begin{eqnarray*}
C_{2, \epsilon_{22}}=&&f_e^2z/2-f_eu_{1e}+\frac{1}{z}[f_e(-u_{t3e}+u_{x3e})\nonumber\\
&&
+u_{1e}^2/2-u_{1e}u_{t2e}],
\end{eqnarray*}
\begin{eqnarray*}
C_{2, \epsilon_{12}}=&&f_nf_ez-(2f_eu_{1n}+u_{1e})+\frac{1}{z}[f_e(-2u_{t3n}\nonumber\\
&&
-2u_{x3e})-u_{t3e}+u_{x3e}-2u_{1e}u_{t2n}\nonumber\\
&&
-2u_1u_{t2e}+u_{1e}(u_1+u_{1n})],
\end{eqnarray*}
\begin{eqnarray*}
C_{2, \epsilon_{11}^2}=&&f_n^2z/4+f_n(-u_{1n}-u_{1e}/2+u_{t2n})\nonumber\\
&&
+u_{1n}^2,
\end{eqnarray*}
\begin{eqnarray*}
C_{2, \epsilon_{22}^2}=f_e^2z/4-f_eu_{1e}/2,
\end{eqnarray*}
\begin{eqnarray*}
C_{2, \epsilon_{12}^2}=&&f_nf_ez/2+f_e(-u_{1n}+u_{t2n})\nonumber\\
&&
-u_{1e}/2+u_{1e}^2/2,
\end{eqnarray*}
\begin{eqnarray*}
C_{2, \epsilon_{11}\epsilon_{12}}=f_nu_{t2e}+2u_{1n}u_{1e},
\end{eqnarray*}
\begin{eqnarray}
C_{2, \epsilon_{11}\epsilon_{22}}=u_{1e}^2/2, C_{2, \epsilon_{12}\epsilon_{22}}=f_eu_{t2e},
\end{eqnarray}
%\end{subequations}

%\begin{subequations}
\begin{eqnarray*}
C_{3, \epsilon_{11}}=&&f_n^2u_1+\frac{1}{z}[f_n^2u_{t3}+f_n(-4u_1^2+2u_1u_{1e}\nonumber\\
&&
+4u_1u_{t2}+2u_{1n}u_{t2}+2u_1u_{t2n})\nonumber\\
&&
+4u_1^2u_{1n}+u_1u_{1e}^2],
\end{eqnarray*}
\begin{eqnarray*}
C_{3, \epsilon_{22}}=&&f_e^2u_1+\frac{1}{z}[f_e^2u_{t3}+f_e(-2u_1u_{1e}+2u_{1e}u_{t2}\nonumber\\
&&
+2u_1u_{t2e})+u_1u_{1e}^2],
\end{eqnarray*}
\begin{eqnarray*}
C_{3, \epsilon_{12}}=&&2f_nf_eu_1+\frac{1}{z}[2f_nf_eu_{t3}+f_e(-4u_1u_{1n}\nonumber\\
&&
+4u_{1n}u_{t2}+4u_1u_{t2n})-2u_1u_{1e}-2u_1u_{1e}^2\nonumber\\
&&
+4u_1^2u_{1e}+2u_{1e}u_{t2}+2u_1u_{t2e}],
\end{eqnarray*}
\begin{eqnarray*}
C_{3, \epsilon_{11}^2}=&&-f_n^3z/2+f_n^2(7u_{1n}/2+2u_{1e}-3u_{t2n}\nonumber\\
&&
-u_{t2e})+f_n(-6u_{1n}^2-u_{1e}^2/2\nonumber\\
&&
-4u_{1n}u_{1e}),
\end{eqnarray*}
\begin{eqnarray*}
C_{3, \epsilon_{22}^2}=&&-f_e^3z/2+f_e^2(u_{1n}/2+2u_{1e}-u_{t2e})\nonumber\\
&&
-f_eu_{1e}^2/2,
\end{eqnarray*}
\begin{eqnarray*}
C_{3, \epsilon_{12}^2}=&&-f_nf_ez/2+f_n^2u_{1e}/2+f_e^2(u_{1n}+u_{1e}/2\nonumber\\
&&
-u_{t2n})
+f_nf_e(3u_1-2u_{t2n}-3u_{t2e})\nonumber\\
&&
+f_e(-2u_{1n}^2-6u_{1n}u_{1e})-3u_{1e}^2/2,
\end{eqnarray*}
\begin{eqnarray*}
C_{3, \epsilon_{11}\epsilon_{12}}=&&-f_n^2f_ez+f_n^2(u_{1e}-2u_{t2e})\nonumber\\
&&
+f_nf_e(4u_{1n}+2u_{1e}-4u_{t2n}-u_{t2e})\nonumber\\
&&
+f_n(-3u_{1e}^2-8u_{1n}u_{1e})\nonumber\\
&&
+f_e(-4u_{1n}^2-2u_{1n}u_{1e}),
\end{eqnarray*}
\begin{eqnarray*}
C_{3, \epsilon_{11}\epsilon_{22}}=-f_nf_eu_{t2e}-2f_eu_{1n}u_{1e}-u_{1e}^2,
\end{eqnarray*}
\begin{eqnarray}
C_{3, \epsilon_{12}\epsilon_{22}}=&&-f_nf_e^2z+f_n^2(2u_{1n}+u_{1e}-2u_{t2n})\nonumber\\
&&
+f_nf_e(2u_{1e}-u_{t2e})\nonumber\\
&&
+f_e(-3u_{1e}^2-2u_{1n}u_{1e}),
\end{eqnarray}
%\end{subequations}

%\begin{subequations}
\begin{eqnarray*}
C_{4, \epsilon_{11}}=&&\frac{1}{z}[f_n^2(2u_1^2-4u_1u_{t2})\nonumber\\
&&
-4f_nu_1^2(u_1+u_{1n})],
\end{eqnarray*}
\begin{eqnarray*}
C_{4, \epsilon_{22}}=\frac{1}{z}[f_e^2(2u_1^2-4u_1u_{t2})-4f_eu_1^2u_{1e}],
\end{eqnarray*}
\begin{eqnarray*}
C_{4, \epsilon_{12}}=&&\frac{1}{z}[f_nf_e(4u_1^2-8u_1u_{t2})-8f_eu_1^2u_{1n}\nonumber\\
&&
-4u_1^2u_{1e}],
\end{eqnarray*}
\begin{eqnarray*}
C_{4, \epsilon_{11}^2}=&&f_n^4z/4+f_n^3(-4u_{1n}-3u_{1e}+3u_{t2n}\nonumber\\
&&
+2u_{t2e})+f_n^2(12u_{1n}^2+7u_{1e}^2/2\nonumber\\
&&
+14u_{1n}u_{1e}),
\end{eqnarray*}
\begin{eqnarray*}
C_{4, \epsilon_{22}^2}=&&f_e^4z/4+f_e^3(-2u_{1n}-3u_{1e}+u_{t2n}\nonumber\\
&&
+2u_{t2e})+f_e^2(7u_{1e}^2/2+2u_{1n}u_{1e}),
\end{eqnarray*}
\begin{eqnarray*}
C_{4, \epsilon_{12}^2}=&&f_n^2f_e^2z+f_n^2f_e(-2u_{1n}-4u_{1e}+u_{t2n}+3u_{t2e})\nonumber\\
&&
+f_nf_e^2(-6u_{1n}-4u_{1e}+5u_{t2n}+3u_{t2e})\nonumber\\
&&
+f_nf_e(4u_{1n}^2+8u_{1e}^2+18u_{1n}u_{1e})\nonumber\\
&&
+f_e^2(6u_1u_{1n})+u_{1e}^2,
\end{eqnarray*}
\begin{eqnarray*}
C_{4, \epsilon_{11}\epsilon_{12}}=&&f_n^3f_ez+f_n^3(-u_{1e}+u_{t2e})+f_n^2f_e(-10u_{1n}\nonumber\\
&&
-5u_{1e}+8u_{t2n}+5u_{t2e})+f_n^2(7u_{1e}^2\nonumber\\
&&
+10u_{1n}u_{1e})+f_nf_e(20u_{1n}^2+5u_{1e}^2\nonumber\\
&&
+22u_{1n}u_{1e}),
\end{eqnarray*}
\begin{eqnarray*}
C_{4, \epsilon_{11}\epsilon_{22}}=&&f_n^2f_e^2z/2+f_n^2f_e(-3u_{1e}+u_{t2e})\nonumber\\
&&
+f_nf_e^2(-2u_{1n}-u_{1e}+2u_{t2n}+u_{t2e})\nonumber\\
&&
+f_nf_e(3u_{1e}^2+6u_{1n}u_{1e})\nonumber\\
&&
+f_e^2(2u_1u_{1n})+u_{1e}^2/2,
\end{eqnarray*}
\begin{eqnarray}
C_{4, \epsilon_{12}\epsilon_{22}}=&&f_nf_e^3z+f_e^3(-2u_{1n}-u_{1e}+2u_{t2n}+u_{t2e})\nonumber\\
&&
+f_nf_e^2(-4u_{1n}-7u_{1e}+2u_{t2n}+5u_{t2e})\nonumber\\
&&
+f_nf_e(5u_{1e}^2+2u_{1n}u_{1e})+f_e^2(4u_{1n}^2\nonumber\\
&&+7u_{1e}^2+14u_{1n}u_{1e}),
\end{eqnarray}
%\end{subequations}

%\begin{subequations}
\begin{eqnarray*}
C_{5, \epsilon_{11}}=\frac{1}{z}(4f_n^2u_1^3),
\end{eqnarray*}
\begin{eqnarray*}
C_{5, \epsilon_{22}}=\frac{1}{z}(4f_e^2u_1^3),
\end{eqnarray*}
\begin{eqnarray*}
C_{5, \epsilon_{12}}=\frac{1}{z}(8f_nf_eu_1^3),
\end{eqnarray*}
\begin{eqnarray*}
C_{5, \epsilon_{11}^2}=&&f_n^4(3u_1/2-u_{t2})-f_n^3u_1(10u_{1n}\nonumber\\
&&
+6u_{1e}),
\end{eqnarray*}
\begin{eqnarray*}
C_{5, \epsilon_{22}^2}=&&f_e^4(3u_1/2-u_{t2})-f_e^3u_1(2u_{1n}\nonumber\\
&&
+6u_{1e}),
\end{eqnarray*}
\begin{eqnarray*}
C_{5, \epsilon_{12}^2}=&&f_n^2f_e^2(6u_1-4u_{t2})-f_n^2f_eu_1(2u_{1n}\nonumber\\
&&
+10u_{1e})-f_nf_e^2u_1(18u_{1n}
+10u_{1e}),
\end{eqnarray*}
\begin{eqnarray*}
C_{5, \epsilon_{11}\epsilon_{12}}=&&f_n^3f_e(6u_1-4u_{t2})
-4f_n^3u_1u_{1e}\nonumber\\
&&
-f_n^2f_eu_1(28u_{1n}+16u_{1e}),
\end{eqnarray*}
\begin{eqnarray*}
C_{5, \epsilon_{11}\epsilon_{22}}=&&f_n^2f_e^2(3u_1-2u_{t2})
-4f_n^2f_eu_1u_{1e}\nonumber\\
&&
-f_nf_e^2u1(18u_{1n}+4u_{1e}),
\end{eqnarray*}
\begin{eqnarray}
C_{5, \epsilon_{12}\epsilon_{22}}=&&f_nf_e^3(6u_1-4u_{t2})-f_e^3u_1(8u_{1n}\nonumber\\
&&
+4u_{1e})-f_nf_e^2u_1(4u_{1n}+16u_{1e}),
\end{eqnarray}
%\end{subequations}

%\begin{subequations}
\begin{eqnarray*}
C_{6, \epsilon_{11}}=0, C_{6, \epsilon_{22}}=0, C_{6, \epsilon_{12}}=0,
\end{eqnarray*}
\begin{eqnarray*}
C_{6, \epsilon_{11}^2}=3f_n^4u_1^2,
\end{eqnarray*}
\begin{eqnarray*}
C_{6, \epsilon_{22}^2}=3f_e^4u_1^2,
\end{eqnarray*}
\begin{eqnarray*}
C_{6, \epsilon_{12}^2}=12f_n^2f_e^2u_1^2,
\end{eqnarray*}
\begin{eqnarray*}
C_{6, \epsilon_{11}\epsilon_{12}}=12f_n^3f_eu_1^2,
\end{eqnarray*}
\begin{eqnarray*}
C_{6, \epsilon_{11}\epsilon_{22}}=6f_n^2f_e^2u_1^2,
\end{eqnarray*}
\begin{eqnarray}
C_{6, \epsilon_{12}\epsilon_{22}}=12f_nf_e^3u_1^2.
\end{eqnarray}
%\end{subequations}
In the above equations, we have employed the following relations,
\begin{equation}
u_{x3e}=u_{3eii}(1-g)^2,
\end{equation}
\begin{equation}
u_{t2n}=u_{2ni}g+u_{2nd},
\end{equation}
\begin{equation}
u_{t2e}=u_{2ei}g+u_{2ed},
\end{equation}
\begin{equation}
u_{t3n}=u_{3nii}g^2+u_{3nid}g+u_{3ndd},
\end{equation}
\begin{equation}
u_{t3e}=u_{3eii}g^2+(u_{3eid1}+u_{3eid2})g+u_{3edd}.
\end{equation}

\bibliographystyle{apsrev4-1}
\bibliography{refs}% Produces the bibliography via BibTeX.

\end{document}